\documentstyle[psbox]{WORKSHOP}
\begin{document}

\title{Swift Follow-up Observations of MAXI Discovered Galactic Transients}

\author{J. A.\ Kennea,$^1$ P. Romano,$^2$ V.\ Mangano,$^2$
 A. P.\ Beardmore,$^3$ P. A.\ Evans,$^3$\\
P. A.\ Curran,$^4$ H. A.\ Krimm,$^5$ K.\ Yamaoka$^6$
\\[12pt]
$^1$  Department of Astronomy \& Astrophysics, The Pennsylvania State  University, 525 Davey Lab, \\University Park, PA 16802, USA\\
$^2$  INAF, Istituto di Astrofisica Spaziale e Fisica Cosmica, Via  U. La Malfa 153, I-90146 Palermo, Italy \\
$^3$  Department of Physics and Astronomy, University of Leicester,  Leicester, LE1 7RH, UK \\
$^4$  AIM, CEA/DSM - CNRS, Irfu/SAP, Centre de Saclay, Bat. 709, FR-91191Gif-sur-Yvette Cedex, France \\
$^5$  Center for Research and Exploration in Space Science and Technology (CRESST),\\ NASA Goddard Space Flight Center, Greenbelt, MD 20771, USA \\
$^6$  Department of Physics \& Mathematics, Aoyama Gakuin University, Sagamihara, Kanagawa 229-8558, Japan \\
{\it E-mail(JAK): kennea@astro.psu.edu} 
}

\abst{ We describe the results of the first year of a program to localize
  new Galactic Transient sources discovered by MAXI with NASA's {\em Swift}
  mission. {\em Swift} is ideally suited for follow-up of MAXI discovered
  transients as its X-ray Telescope (XRT) field of view ($\sim0.2$ degrees
  radius) is closely matched to the typical MAXI error circle. The XRT is
  capable of localizing new sources to an accuracy of up to 1.5 arc-seconds
  radius (90\% confidence), and the {\em Swift} Optical/UV Telescope also
  provides optical imaging of any optical counterpart of the X-ray
  source. If no optical counterpart is found with {\em Swift} (usually due
  to absorption), the XRT position is good enough to allow for ground based
  IR telescopes to positively identify the optical counterpart. Although
  localization and identification of MAXI transients is the main aim of the
  program, these are often followed up by long term monitoring of the
  source. We present here results from 2 of these monitoring programs: the
  black-hole candidate MAXI J1659$-$152, and the Be/X-ray binary candidate
  MAXI J1409$-$619.  }

\kword{X-rays: individual: LS V$+4$4 17,    HD 347929/1RXS J180724.2$+$194217,   
SAX J1452.8$-$5949,  MAXI J1659$-$152, MAXI J1409$-$619, 4U 1137$-$65/GT Mus }

\maketitle
\thispagestyle{empty}

\section{Introduction}

The process of accretion that drives most X-ray astrophysical phenomena can
often be dramatic and short lived, with increases in accretion rates
causing X-ray flux increases of up to 6 orders of magnitude from quiescent
levels. In many cases these events lead to the discovery of previously
unknown systems, or systems that were previously considered
uninteresting. The flare-up of a Galactic X-ray transient typically heralds
a rapid increase in the rate of accretion onto a compact object (white
dwarf, neutron star or black hole), and provides an ideal laboratory for
studying astrophysics in a relativistic regime. Even though transients have
been studied for many years, our understanding of the processes behind
extreme accretion events remains relatively poor. X-ray transients also
cover a wide range of different system phenomenology, including Black Hole
binaries systems (e.g.\ Remillard \& McClintock 2006), Low Mass X-ray
Binaries, High Mass X-ray Binaries (HMXB), millisecond pulsars (e.g.\
Campana et al.\ 2008), Cataclysmic Variables, Novae and Supergiant Fast
X-ray Transients (e.g.\ Romano et al.\ 2010).

These transient events are rare, and often short lived, making detection
and detailed study difficult. To obtain a good rate of detection of
transient outbursts, X-ray instruments that cover very large areas of the
sky are required. However, these wide field instruments typically lack the
spatial resolution required to provide accurate localizations necessary for
further optical and IR observations, and often do not have enough spectral
sensitivity for a detailed analysis of the characteristics of the
outburst. Therefore follow-up observations with complementary
observatories, such as {\em Swift}, are required to provide more accurate
positions and simultaneously observe the broadband (Optical to Gamma-ray)
spectral behavior.

``Monitor of All-sky X-ray Image'' (MAXI, e.g.\ Ueno et al.\ 2009), part of
the Japanese Experiment Module on the International Space Station, provides
a powerful tool for discovery of new X-ray transients. MAXI's ability to
perform a near all-sky X-ray image of the sky in the 0.5--20 keV energy
band, with sensitivities as low as 60 mCrab (5 sigma) in a single orbit and
15 mCrab in a day, makes it more capable of finding transients than other
instruments like {\em Swift}'s Burst Alert Telescope (BAT; Barthelmy et al.\
2004) which has too hard an energy band, or RXTE PCA Galactic Bulge scans
(limited spatial and temporal coverage).9 MAXI's capability of finding
transient X-ray phenomena has been proven in the first year of operations
by the detection of outbursts of known sources, GRBs, and most recently the
discovery of a new Galactic transients: MAXI J1659$-$152 and MAXI
J1409$-$619.

We report here on the results of a program to localize MAXI discovered
X-ray Galactic X--ray transients utilizing NASA's {\em Swift} Gamma-Ray
Burst Explorer Mission (e.g.\ Gehrels et al.\ 2004). {\em Swift} has proven
capabilities in the follow-up of transient sources: fast slewing, flexible
scheduling, and the ability to perform rapid TOO observations of targets
through spacecraft commanding as quickly as within 1 hour of the
announcement of a target's location. Recent developments of the {\em Swift}
planning infrastructure have been driven towards making these observations
both quicker and less burdensome on the small {\em Swift} mission
operations team. {\em Swift}'s X-ray Telescope (XRT; Burrows et al.\ 2004)
has a field of view of approximately $23.6'$ diameter, well matched to the
typical $0.2$ degree error circle from MAXI. The {\em Swift} UV/Optical
Telescope (UVOT; Roming et al.\ 2005) here also provides two valuable
services in support of these observations: broad band optical/UV
observations of the optical counterpart of the new transient (if visible);
and astrometric correction of XRT data that allows X-ray localization
errors to be reduced to as little as 1.5 arc-seconds radius (90\%
confidence).

\begin{figure*}[t]
\centering
\psbox[xsize=12cm]
{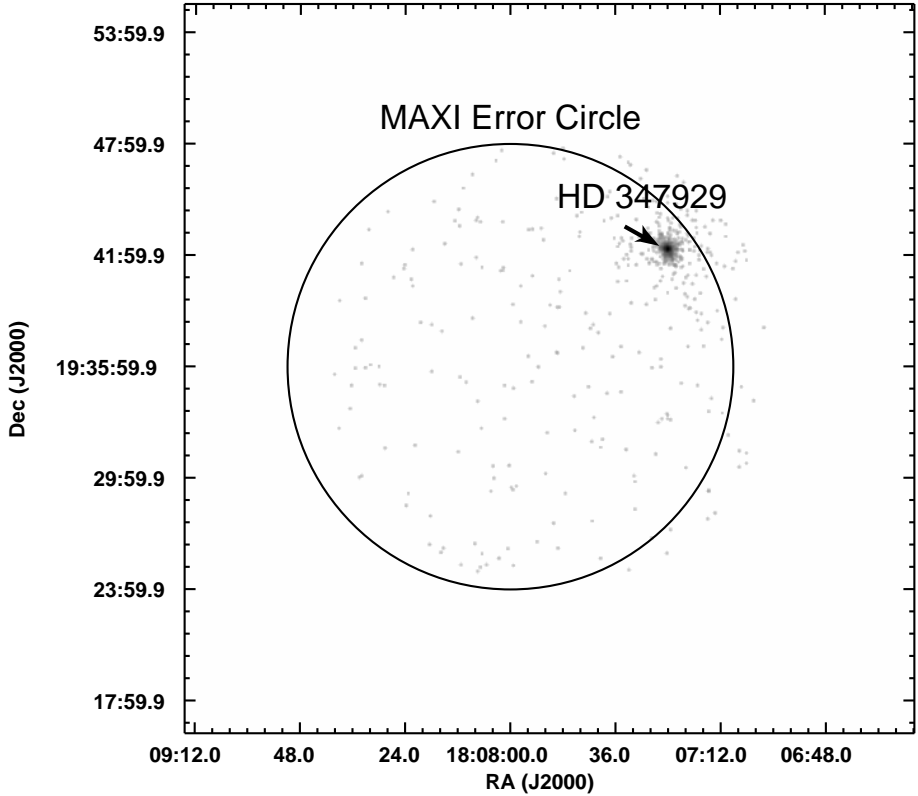}
\caption{\label{fi:HD347929local}The XRT field of view of HD 347929,
  with MAXI transient error circle marked. This observation is typical of
  what is seen in this program for other follow-ups.}
\end{figure*}

\section{Observation Planning and Analysis}

Observations of new transients by {\em Swift} are triggered from results of the
MAXI Nova Alert System (Negoro et al.\ 2009), which reports on outbursts
from targets on both single orbit timescales and longer. Decision to
trigger {\em Swift} observations is made based on the following criteria: Is the
source not obviously identified with a already known X-ray transient? Is
the MAXI error circle small enough to be covered adequately by the XRT
field of view? Is the new transient Galactic in origin? Is the new
transient sufficiently bright and long-lived to still be visible by
{\em Swift}?

Triggers that meet this criterion are submitted to the {\em Swift} Mission
Operations Center (MOC) as a high priority Target of Opportunity (TOO)
request for observations within 24 hours. {\em Swift} observations
typically occur within 24 hours of the MAXI trigger, but can be as quick as
within an hour of the TOO request being submitted. 

{\em Swift}'s low background means that a credible detection and
localization can be made for sources in 1ks with just 10 counts. XRT's
sensitivity corresponds to a count rate of approximately 0.7 c s$^{-1}$
mCrab$^{-1}$. Given that MAXI transients typically have brightness of 40
mCrab and above this makes it trivial to detect such transients with {\em
  Swift}, unless they have rapidly faded. UVOT obtains exposures in B,
V and White filters, which provide the best chance of being able to utilize
UVOT data to perform astrometric corrections. XRT is forced into Photon
Counting mode to ensure imaging mode data is taken.

Analysis of {\em Swift} data is performed utilizing the methods presented
in Evans et al.\ (2009), via an automated process. The XRT point source is
localized using PSF fitting, and then astrometric correction to reduce
systematic errors. In the case of heavy pile-up, PSFs with modelled pile-up
are used to fit the position and the core of the PSF is excluded to remove
the effect of pile-up from the extracted light-curves and spectra of the
point source.  The results of this analysis are published rapidly via The
Astronomers Telegram\footnote{\tt http://www.astronomerstelegram.org}, typically
within 24 hours of the observation.

\begin{figure*}[t]
\centering
\psbox[xsize=17cm,rotate=r]
{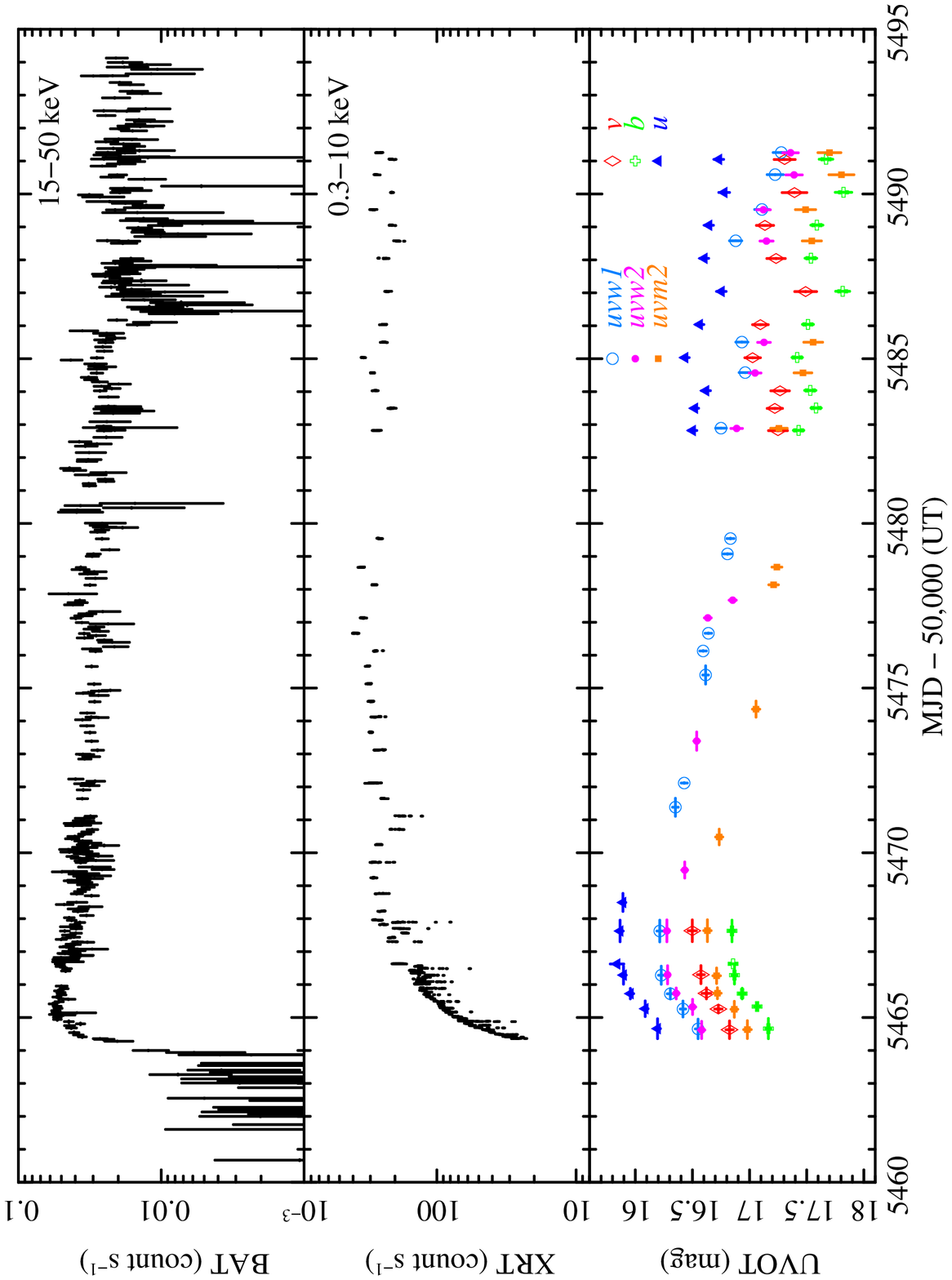}
\caption{\label{fi:maxij1659lc}The {\em Swift} light-curves of MAXI
  J1659$-$152 during its visibility period of 27 days starting September
  25th, 2010. Presented here are results from the BAT Transient Monitor,
  the XRT count rate, and the UVOT light-curve in multiple filters (as
  labeled).}
\end{figure*}

\section{Results}

In this section we summarize the results of the observations performed as
part of this program covering the time period of April 2010 to the end of
December 2010. We provide localizations, identifications, source types and
details of any {\em Swift} monitoring of the objects where applicable.  During
2010 the program was triggered on 6 occasions. Results are presented in
chronological order of discovery.

\subsection{LS V +44 17}

On March 31st, 2010 at 02:10\,UT, MAXI detected a transient outburst (Morii
et al.\ 2010), which was reported to be consistent with the location of
LS~V~+44~17, a Be/X-ray binary system. A {\em Swift} TOO was submitted to
follow-up and confirm the identification of this transient. However before
a TOO could be uploaded at 18:34\,UT on April 1st, 2010 BAT triggered on
LS~V~+44~17 and {\em Swift} observed this source with the XRT and UVOT
autonomously. An XRT localization confirmed that the X-ray transient was
indeed an outburst of LS~V~+44~17 (Stratta et al.\ 2010).

\subsection{HD 347929/1RXS J180724.2+194217}

On June 27th, 2010  at 08:27\,UT MAXI detected an X-ray transient. Usui et
al. (2010) reported that the source was likely an outburst of the RS CVn
system HD 347929 (also named 1RXS J180724.2+194217). A {\em Swift} follow-up TOO was
performed to confirm the identification at 00:10\,UT on June 29th,
2010. {\em Swift}/XRT detected a bright X-ray source that was consistent with the
coordinates for HD 347929, confirming that the MAXI transient was indeed an
outburst of the RS CVn system (Kennea et al.\ 2010a). 
Figure~\ref{fi:HD347929local} shows the {\em Swift}/XRT field on HD
347929, and the MAXI error box, as an example of a typical follow-up of a
0.2 degree MAXI transient error circle.

\subsection{SAX J1452.8-5949}

A MAXI discovered X-ray transient consistent with the location of SAX
J1452.8$-$5949 was reported (Kawai, private communication) on August 17th,
2010. MAXI detected the source to be approximately 100 times brighter than
seen in an archival XMM-Newton observation of the source.

Two {\em Swift} observations were performed, one centered on the MAXI error circle,
and one centered on SAX J1452.8$-$5949. The observations show a weak
detected source at the position of SAX J1452.8$-$5949, but no evidence of
enhanced emission. 

\begin{figure*}[t]
\centering
\psbox[xsize=17cm,rotate=r]
{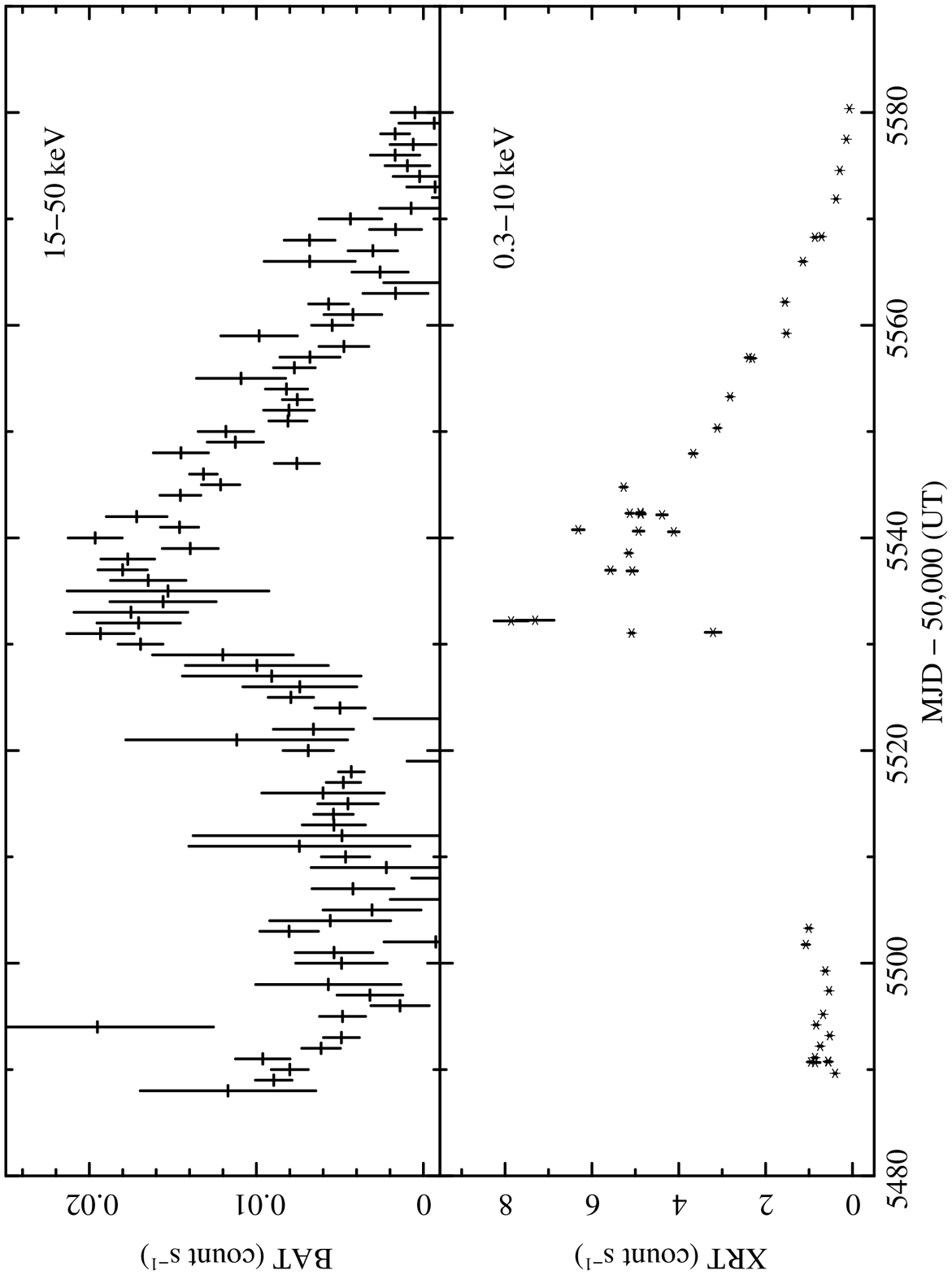}
\caption{\label{fi:maxij1409lc}The {\em Swift} light-curves of MAXI
  J1409$-$619 between October 20th, 2010 and January 19th, 2011. Shown here
  are the BAT transient monitor light-curves and the XRT 0.3--10 keV
  light-curve. Note that the gap in the XRT lightcurve between November
  3rd, 2010 and November 30th, 2010 caused by MAXI J1409$-$619 being
  unobservable due to being too close to the Sun.}
\end{figure*}

\subsection{MAXI J1659$-$152}

MAXI J1659$-$152 was first reported when it triggered the {\em Swift}/BAT
at 08:05\,UT on September 25th, 2010 (Mangano et al.\ 2010), but was
misidentified as a Gamma-Ray Burst and named GRB 101225A. Follow-up
observations by the {\em Swift}/XRT and UVOT occurred 31 mins after the BAT
detection and localized the bright transient, finding it to be a new
uncatalogued X-ray source. Negoro et al.\ (2010) reported that MAXI had
detected the source at 02:30\,UT, approximately 5.5 hours before the BAT
detection, confirming that this was not a GRB, but in fact a previously
unknown X-ray transient.

The XRT enhanced localization of the source found the following
coordinates: RA, Dec = $16^h59^m01^s.56$, $-15^\circ15'30''.5$ (J2000,
$1.7''$ error radius, 90\% confidence). A UVOT counterpart close to the XRT
error circle was found at the following coordinates: RA, Dec =
$16^h59^m01^s.679$, $-15^\circ15'28''.54$ (J2000; $0.70''$ error radius,
90\% confidence). This UVOT error circle was consistent with a later EVLA
radio detection of the transient (Paragi et al.\ 2010), and therefore is
considered the best {\em Swift} localization of the transient. Observed
correlated variability of the UVOT counterpart also unambiguously confirms
its association with MAXI J1659$-$162.

{\em Swift} performed follow-up monitoring observations of MAXI J1659$-$152 for
approximately 27 days following its initial detection, after which the
source was no longer visible due to a Sun constraint. In
Figure~\ref{fi:maxij1659lc} we show the BAT, XRT and UVOT light-curves of
the transient, along with the XRT hardness ratio. The transient shows
considerable broadband spectral evolution. The 15--150 keV BAT flux shows a
standard fast rise/exponential decay shape, similar to many X-ray
transients. The XRT rise was slower to peak, and slower to decay. The UVOT data
showed similar rise to peak, followed by a long decay. 

The XRT+BAT combined spectrum of MAXI J1659$-$152 is well fit by a model
that consists of an absorbed power-law + disk blackbody spectrum, which is
typical for blackhole binary systems. MAXI J1659$-$152 also goes through
several canonical state changes (for an explanation see Remillard and
McClintock, 2006), which are typically seen in blackhole binary systems. In
observations taken during the initial detection the spectrum can be well
described by a hard power-law ($\Gamma \simeq 1.6$), typical of a blackhole
binary system in the Hard State. The source rapidly softens to the Steep
Power-Law State ($\Gamma \simeq 2.5$) around September 27th,
2010. Following this the spectrum becomes increasingly dominated by a $kT
\simeq 1$ keV disk blackbody component, signifying the evolution of the
source into the Thermal State, although this state transition never
occurred by the end of the {\em Swift} observation period.

QPOs were observed in the source by both {\em Swift} and RXTE (Kalamar et al.\
2010), a further signature that MAXI J1659$-$152 is a blackhole
candidate. Furthermore the {\em Swift}/XRT light-curve shows significant and
variable dips in the X-ray light-curve, suggestive of eclipsing and/or
dipping due to absorption by the accretion disk itself. Analysis of the XRT
light-curve using a Lomb-Scargle technique (Scargle et al.\ 1982) finds a
period of 2.42$\pm$0.09 hours, which is presumed to be the orbit
period. This periodicity was also observed by RXTE (Belloni et al.\ 2010)
and XMM-Newton (Kuulkers et al.\ 2010) observations. If this periodicity is
the orbital period, this would make MAXI J1659$-$152 the shortest period
blackhole binary system yet discovered.

A detailed analysis of the {\em Swift} data from MAXI J1659$-$152 is given by Kennea
et al.\ (2011).

\subsection{MAXI J1409$-$619}   

On October 17th, 2010 MAXI reported the detection of MAXI J1409$-$619 at a
level of 41 mCrab (Yamaoka et al.\ 2010). {\em Swift} observed the transient at
15:14\,UT, on October 20th, 2010 and found a bright, uncatalogued point
source at the following coordinates: RA, Dec = $14^h08^m02.56^s$, 
$-61^\circ59'00''.3$ (J2000, $1.9''$ error radius, 90\% confidence),
identifying the source as a previously unknown X-ray transient (Kennea et
al. 2010b).

{\em Swift} monitored MAXI J1409$-$619 for 23 days until it went into a Sun
constraint, during that period the source showed a variable light-curve (see
Figure~\ref{fi:maxij1409lc}) ranging between 0.5--1 XRT count s$^{-1}$. The
X-ray spectrum was well described by an absorbed power-law model, with a
high absorption ($N_{\rm H} = 4 \times 10^{22}$ cm$^{-2}$). 

On November 30th, 2010 at 15:35\,UT, after the source had come out of a
{\em Swift} Sun constraint, BAT triggered on MAXI J1409$-$619 and found the source
to be in a higher state, approximately 7 times brighter than seen in the
previous observations. Furthermore, analysis of the XRT light-curve showed
the presence of an $\sim500$ second periodicity, which was not detected in the earlier
data (Kennea et al.\ 2010c). This periodicity likely originates from the
neutron star compact object in the system, and as such is the pulsar period
of MAXI J1409$-$619. This period was confirmed by RXTE (Yamamoto et al.\ 2010)
and Fermi/GBM (Camero-Arranz et al.\ 2010). The presence of the $\sim500$ second
periodicity strongly indicates that MAXI J1409$-$619 is HMXB source, with the
transient nature making it likely to be a Be/X-ray binary system, although
the source type has yet to be confirmed. 

Monitoring observations of MAXI J1409$-$619 performed by XRT show that
the source continued to brighten to peak around November 30th, 2010,
followed by a decay into quiescence in late January, 2011 (Figure~\ref{fi:maxij1409lc}).

\subsection{4U 1137-65/GT Mus}

MAXI reported an outburst of a transient whose localization is consistent
with the RS CVn star GT Mus, a.k.a. 4U 1137$-$65 (Nakajima et al.\ 2010) on
November 10th, 2010 at 6:17\,UT. {\em Swift} performed a target of opportunity
observation of the MAXI error circle at 23:13\,UT for 1ks. {\em Swift} detected a
bright X-ray source whose localization is consistent with that of GT Mus,
confirming that the MAXI detection was indeed an outburst of this known
source (Kennea et al.\ 2010d).

\section{Conclusions}

We have presented the results of a {\em Swift} program to rapidly follow-up
and localize MAXI discovered X-ray transients. Typically, the MAXI
localization of a source ($\sim 0.2$ degrees) is well matched to the XRT
field of view, making XRT the ideal instrument to attempt to more
accurately localize the transient. {\em Swift}'s unique observing
flexibility also allows us to rapidly follow-up and report on the detection
of these transients, usually within 24 hours of being notified of the
transient by the MAXI team. {\em Swift} is capable of localizing transients
to an accuracy of up to 1.5 arc-seconds radius (90\% confidence), or even
more accurate if a UVOT counterpart is found.

We have reported on 6 triggers of the {\em Swift}/MAXI transient follow-up
program. Two of those triggers localized and confirmed the present of a
previously unknown transient source, one (MAXI J1659$-$152) is a new
blackhole binary system. In both cases {\em Swift} follow-up and monitoring
observations of these targets was triggered, giving valuable insight into
the nature of these transients. For the other 3 triggers {\em Swift} was
able to positively identify the source of the MAXI trigger. In only 1 case
was the result of MAXI follow-up inconclusive (SAX J1542.8$-$5949). 

The unique complementary capabilities of {\em Swift} and MAXI has proven to
be well matched in the search for and localization of new Galactic
transients sources. The all-sky monitoring of MAXI, combined with {\em
  Swift}'s rapid response TOO capability and the convenient similarity of
the XRT's field of view with the typical MAXI error circle, has proved to
be very successful in accurately localizing new transient sources, which
provides an important service to the community.

\section{Acknowledgements}
This work is support by NASA grant NNX10AK40G, through the Swift Guest
Investigator Program.  PR and VM acknowledge financial contribution from
the agreement ASI-INAF I/009/10/0. This work made use of data supplied by
the UK Swift Science Data Centre at the University of Leicester. We
acknowledge the use of public data from the Swift data archive.

\section*{References}

\re
Barthelmy, S. D., et al. 2005, Space Sci. Rev., 120, 143

\re
Belloni, T. M., Motta, S., Mun\~oz-Darias, T. 2010, The Astronomer's Telegram, 2926

\re
Burrows, D. N., et al. 2005, Space Sci. Rev., 120, 165

\re
Camero-Arranz, A., Finger, M. H.,  Jenke, P., 2010, The Astronomer's
Telegram, 3069

\re
Campana, Stella and Kennea, ``{\em Swift} Observations of SAX J1808.4$-$3658: Monitoring the Return to
Quiescence'', 2008, ApJ, 684, 99

\re
Evans, P. A., et al. 2009, MNRAS, 397, 1177

\re
Gehrels, N., et al. 2004, ApJ, 611, 1005

\re
Kalamar, M., et al. 2010, The Astronomer's Telegram, 2881

\re
Kennea, J. A., et al. 2010a, The Astronomer's Telegram, 2701

\re
Kennea, J. A., et al. 2010b, The Astronomer's Telegram, 2962

\re
Kennea, J. A., et al. 2010c, The Astronomer's Telegram, 3060

\re
Kennea, J. A., et al, 2010d, The Astronomer's Telegram, 3025

\re
Kennea, J. A., et al, 2011, ApJ, {\em In press}

\re
Kuulkers, E., et al. 2010, The Astronomer's Telegram, 2912

\re
Mangano, V., et al.\ 2010, GCN, 11296

\re
Nakajima, M., et al.\ 2010, The Astronomer's Telegram, 3021
\re
Negoro, H. 2009, Astrophysics with All-Sky X-Ray Observations,
60

\re
Paragi, J. A., et al, 2010, The Astronomer's Telegram, 2906

\re
Remillard, R. A., \& McClintock, J. E. 2006, ARA\&A, 44, 49

\re
Romano, La Parola, et al.\ ``Two years of monitoring Supergiant Fast X-ray Transients with {\em Swift}'',
2011, MNRAS, 410, 1825  

\re
Roming, P. W. A., et al. 2005, Space Sci. Rev., 120, 95

\re
Scargle, J. D. 1982, ApJ, 263, 835

\re
Stratta, G., et al.\ 2010, GCN, 10561

\re
Ueno et al.\ ``The MAXI Mission Overview and Schedule'', Astrophysics with All-Sky X-ray
Observations, Proceedings of the RIKEN symposium, p. 8.

\re
Usui et al.\ 2010, The Astronomer's Telegram, 2700

\end{document}